\documentclass[a4paper,10pt,twoside]{cpc-hepnp}

\usepackage{multicol}
\usepackage{graphicx}
\usepackage{booktabs}
\usepackage{amssymb,bm,mathrsfs,bbm,amscd}
\usepackage[tbtags]{amsmath}
\usepackage{lastpage}
\usepackage{CJK}
\usepackage{subfigure}

\begin{document}

\fancyhead[c]{\small Submitted to Chinese Physics C}
\fancyfoot[C]{\small 010201-\thepage}

\footnotetext[0]{Received \number\day ~October 2015}

\title{Numerical Modeling of Thermal Loading of Diamond Crystal in X-ray FEL Oscillators \thanks{Supported by National Natural Science Foundation of China (11175240, 11205234, 11322550) and Program for Changjiang Scholars and Innovative Research Team in University (IRT1280)} }

\author{%
      SONG Mei-Qi $^{1}$ 
\quad ZHANG Qing-Min $^{1,1)}$ \email{zhangqingmin@mail.xjtu.edu.cn}%
\quad GUO Yu-Hang $^{1}$ \\
\quad LI Kai $^{2}$
\quad DENG Hai-Xiao $^{2}$
}
\maketitle

\address{%
$^1$ Department of Nuclear Science and Technology, School of Energy and Power Engineering, \\
Xi¡¯an Jiaotong University, Xi¡¯an 710049, China\\
$^2$ Shanghai Institute of Applied Physics, Chinese Academy of Sciences, Shanghai 201800, China\\
}

\begin{abstract}
Due to high reflectivity and high resolution of X-ray pulses, diamond is one of the most popular Bragg crystals serving as the reflecting mirror and mono$-$chromator in the next generation of free electrons lasers (FELs). The energy deposition of X-rays will result in thermal heating, and thus lattice expansion of the diamond crystal, which may degrade the performance of X-ray FELs. In this paper, the thermal loading effect of diamond crystal for X-ray FEL oscillators has been systematically studied by combined simulation with Geant4 and ANSYS, and its dependence on the environmental temperature, crystal size, X-ray pulse repetition rate and pulse energy are presented. Our results show that taking the thermal loading effects into account, X-ray FEL oscillators are still robust and promising with an optimized design.
\end{abstract}




\footnotetext[0]{\hspace*{-3mm}\raisebox{0.3ex}{$\scriptstyle\copyright$}2013
Chinese Physical Society and the Institute of High Energy Physics
of the Chinese Academy of Sciences and the Institute
of Modern Physics of the Chinese Academy of Sciences and IOP Publishing Ltd.}%

\begin{multicols}{2}

\section{Introduction}

Driven by the successful operation of hard X-ray free-electron lasers (FELs) \cite{lab1,lab2} and the growing interest of scientific users, several hard X-ray FEL facilities are being constructed and designed around the world. Currently, all these FELs use self-amplified spontaneous emission (SASE) \cite{lab3} as the lasing mode for the hard X-rays, which starts from the beam noise, and results in radiation with transverse coherence, but poor longitudinal coherence. Therefore, various seeded FEL schemes have been intensively studied \cite{lab4,lab5,lab6} worldwide to generate fully coherent FEL radiation, especially in the soft X-ray regime.

In the hard X-ray regime, with the development of high-reflectivity and high-resolution crystals, alternative fully coherent FEL schemes are proposed and are of great interest. One is the self-seeding approach, which has been experimentally demonstrated to narrow the SASE bandwidth \cite{lab7,lab8}. In this scheme, the noisy SASE radiation generated in the first undulator section is spectrally purified by a crystal filter. Then, in the second undulator section, this spectrally purified FEL pulse serves as a highly coherent seed to interact with the electron bunch again to configure a FEL amplifier, which could significantly improve the temporal coherence of the final output radiation. The other scheme is a low-gain oscillator which has been reconsidered as a promising candidate for hard X-ray FELs through the use of an X-ray crystal cavity, known as XFELO \cite{lab9,lab10,lab11,lab12}. It is widely believed that with the peak brilliance comparable to SASE and the average brilliance several orders of magnitude higher than SASE, XFELO may open up new scientific opportunities in various research fields.

In these next generation FELs, however, the spectrum bandwidth of the X-rays is heavily dependent on the quality of the crystal. That is to say, the interaction between the X-ray FEL pulse and the crystal can cause energy deposition in the crystal and result in the raising of the crystal temperature, further expanding the crystal volume and lattice. Hence, the Bragg energy is shifted and the reflection spectrum enlarged, which is called the thermal loading effect of X-rays and is not negligible because of the high brightness of the X-ray pulse. In this paper, we numerically investigate the interaction between X-ray pulses and diamond crystals in XFELO by using Geant4 \cite{lab13} and its thermal conduction through ANSYS \cite{lab14}. In section 2, the numerical model, the general result of energy deposition and temperature rise are described. The thermal loading dependence on the environmental temperature, crystal size, X-ray repetition rate and pulse energy are illustrated and discussed in section 3. Finally, we summarize the optimal parameters of the diamond crystal and X-ray pulse in the conclusions.

\section{Numerical  Modeling}

To study the thermal loading effects, the interactions between the incident X-ray photon pulse and diamond crystal are first modeled using Geant4 to get the photon energy deposition and thermal conversion. Then the collected data are imported to ANSYS to investigate the temperature shift of the crystal during penetration of the X-ray pulse.

According to the early theoretical results of XFELO \cite{lab8,lab9,lab10,lab11,lab12}, a set of baseline parameters are assumed for the X-ray pulses here. Due to the narrow spectrum bandwidth of XFELO, we take the X-ray wavelength as constant and assume the photon energy is 12.04 keV, with $n=1\times 10^{10}$ photons per pulse and 1 MHz repetition rate. In the longitudinal dimension, the X-ray pulse is assumed to be uniformly distributed within a time window of $\tau = 0.5$ ps, while it has a Gaussian distribution with RMS size 65 $\mu$m in the transverse direction. For the diamond crystal, the considered interaction transverse region is 1 mm$\times$1 mm, and the thickness is 200 $\mu$m.

In order to simplify the description, we define a coordinate system (x,~y,~z), with x-axis and y-axis parallel to the crystal surface while the z-axis is perpendicular to it. We suppose the X-ray photon pulse propagates along the z-axis. In the Geant4 setup, outside the detector is air (2 mm$\times$2 mm$\times$0.6 mm), while the inside is carbon atoms which set as the sensitive detector (1 mm$\times$1 mm$\times$0.2 mm). When X-ray energy deposition happens in the sensitive detector, the deposition location and amplitude are recorded. In our simulation, the X-ray intensity is acquired by applying Gaussian sampling along the x and y axes. As the incident photon energy is 12.04 keV, which is far less than the threshold 1.02 MeV of the electron pair effect, only the photoelectric effect and Compton scattering effects are involved.

The X-ray energy deposition distribution function $\triangle$E(x,y,z) can be obtained by summing up the energy deposition in the given small volume $\triangle V$. Then the converted thermal power distribution during the duration of the X-ray pulse is given by
\begin{equation}
\label{eqn1}
h(x,y,z,t)=\frac{\triangle E(x,y,z)}{\tau \triangle V} \frac{n}{N_e}(1-R)
\end{equation}
where $N_e$ is the number of the total simulated events in Geant4, and R is the X-ray reflectivity of diamond crystal.

If we consider an X-ray pulse propagating through a 50 mm-thick crystal which is sufficient to absorb almost all photons, as demonstrated in Fig.~\ref{de}, the energy deposition will decrease exponentially with the depth of crystal. In further studies with the secondary electrons taken into account, an increase during the initial hundreds of nanometers exists \cite{lab15,lab16}, which is invisible here, because this is significantly small compared to the thickness of the crystal. In practice, the thickness of diamond crystal is usually hundreds of micrometers (we take a value of 200 $\mu$m), thus the X-ray energy deposition dependence on the crystal thickness can be ignored, as shown in the inset of Fig.~\ref{de}.
\begin{center}
\includegraphics[width=8cm]{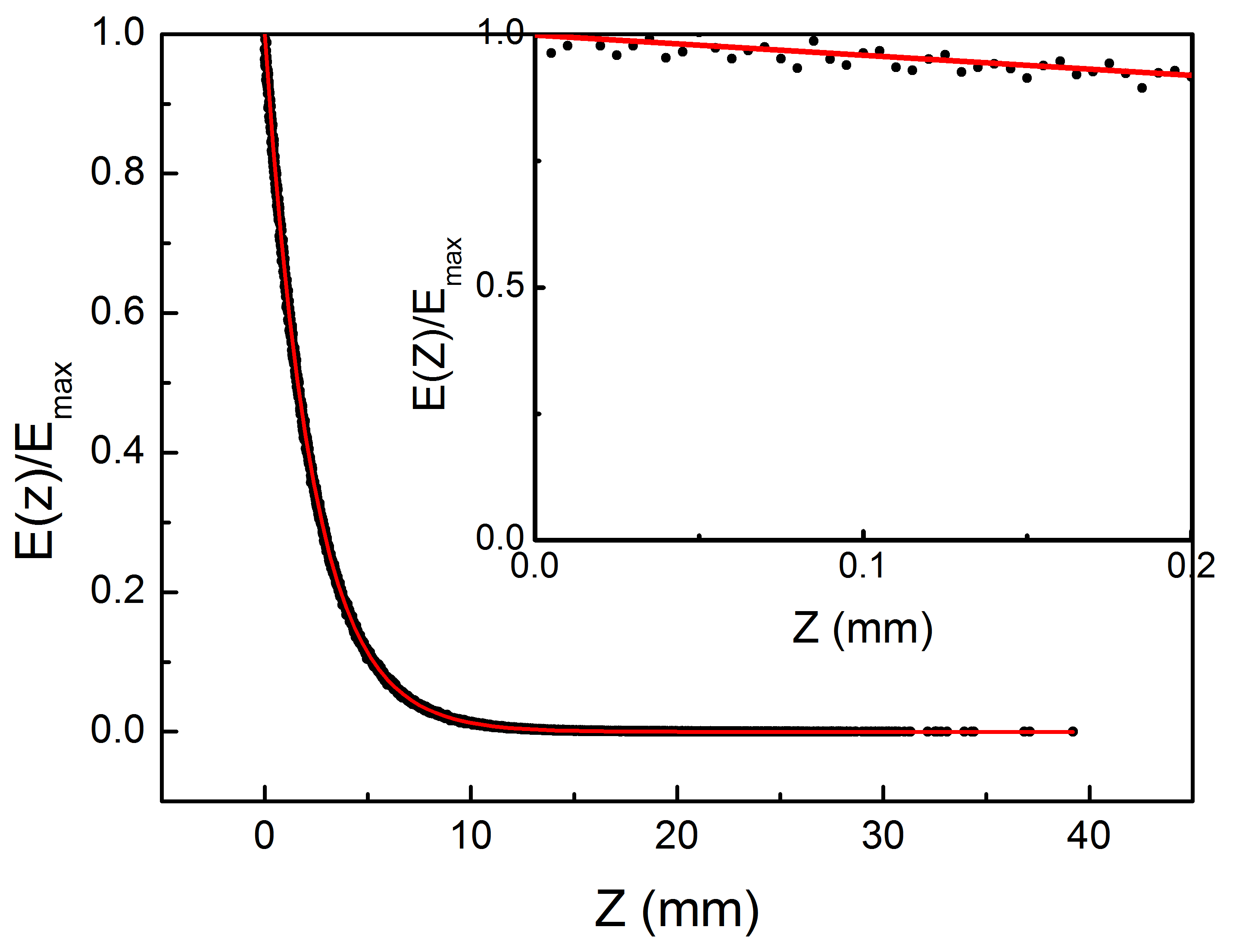}
\figcaption{\label{de}   X-ray energy deposition along the longitudinal direction within the crystal. (The energy deposition is normalized) }
\end{center}

Fig.~\ref{distribution} shows the transverse X-ray energy deposition. When secondary electrons are ignored, it has a Gaussian distribution, which resembles the map of the power distribution of the incident X-ray pulse.
\begin{center}
\includegraphics[width=8cm]{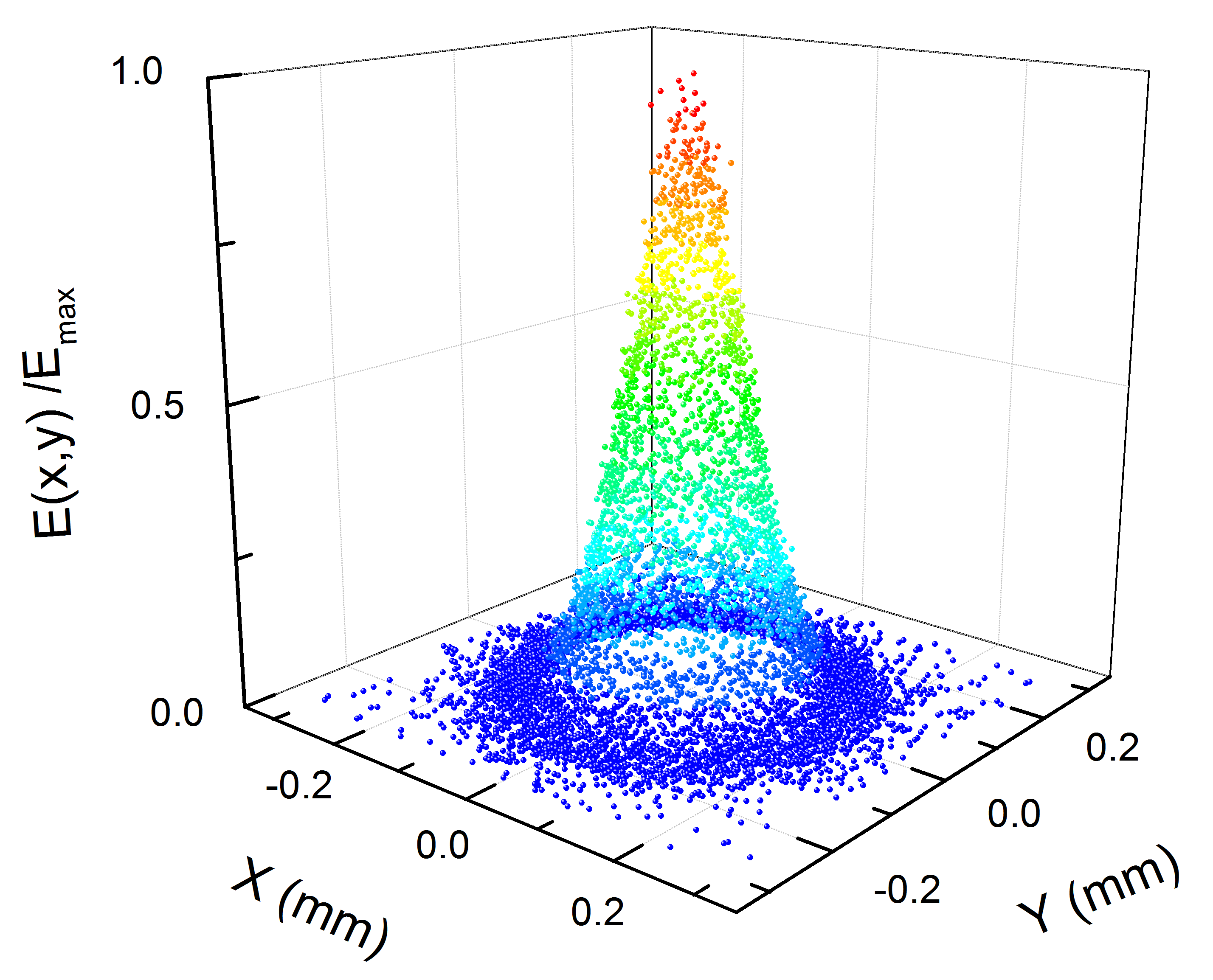}
\figcaption{\label{distribution}   X-ray energy deposition along the transverse direction of the crystal. (The energy deposition is normalized)}
\end{center}

When an X-ray pulse containing 1$\times 10^{10}$ photons (nearly 19.3 $\mu$J) is injected into a diamond crystal with reflectivity of 90\%, the thermal power distribution derived from simulation is written as
\end{multicols}
\begin{equation}
\label{eqn2}
h(x,y,z,t)=\left\{\begin{array}{lcl} 5.296\times 10^{16} e^{-\frac{x^2+y^2}{8.521\times 10^{-9}}}e^{-436z} &~~& (10^{-6}k\leq t \leq 10^{-6}k+5\times 10^{-13})\\
  0 &~~& \left( 10^{-6}k+5\times 10^{-13} \leq t \leq 10^{-6}(k+1)\right) \end{array}\right.
\end{equation}
\begin{multicols}{2}
where k=0,1,2,$\cdots$. x,y and z are in units of m, h and t are in W/m$^3$ and second respectively. According to Eq.~(\ref{eqn2}), nearly 0.15 $\mu$J of X-ray photon energy will be absorbed in a 200 $\mu$m thick crystal.

In order to acquire the thermal behavior of diamond crystal under exposure to X-ray pulses, a heat source according to the thermal power distribution in Eq.~(\ref{eqn2}) is imported into ANSYS to study the thermal transportation of the diamond crystal. In the ANSYS transient thermal analysis module, grid size and time step size are optimized to save computer memory and calculation time, while ensuring accuracy. For the 0.5 ps long X-ray pulse of 1 MHz repetition rate in our study, time steps of 1$\times 10^{-13}$ s and 2$\times 10^{-8}$ s are chosen for the temperature rise and heat dissipation process, respectively.

Under an environmental temperature of 300 K, the temperature distribution at different points of the crystal was simulated. For a thin crystal, the hottest location is the center of the crystal due to thermal conduction, which is the most important point that we should focus on in the following discussion. Fig.~\ref{temperature300k} presents one typical temperature change at the center of the crystal. Temperature change has two components $\triangle T_1$ and $\triangle T_2$, which are caused by short-term heating due to X-ray photon energy deposition and long-term heating accumulation of remaining heat in the crystal, respectively.
\begin{center}
\includegraphics[width=8cm]{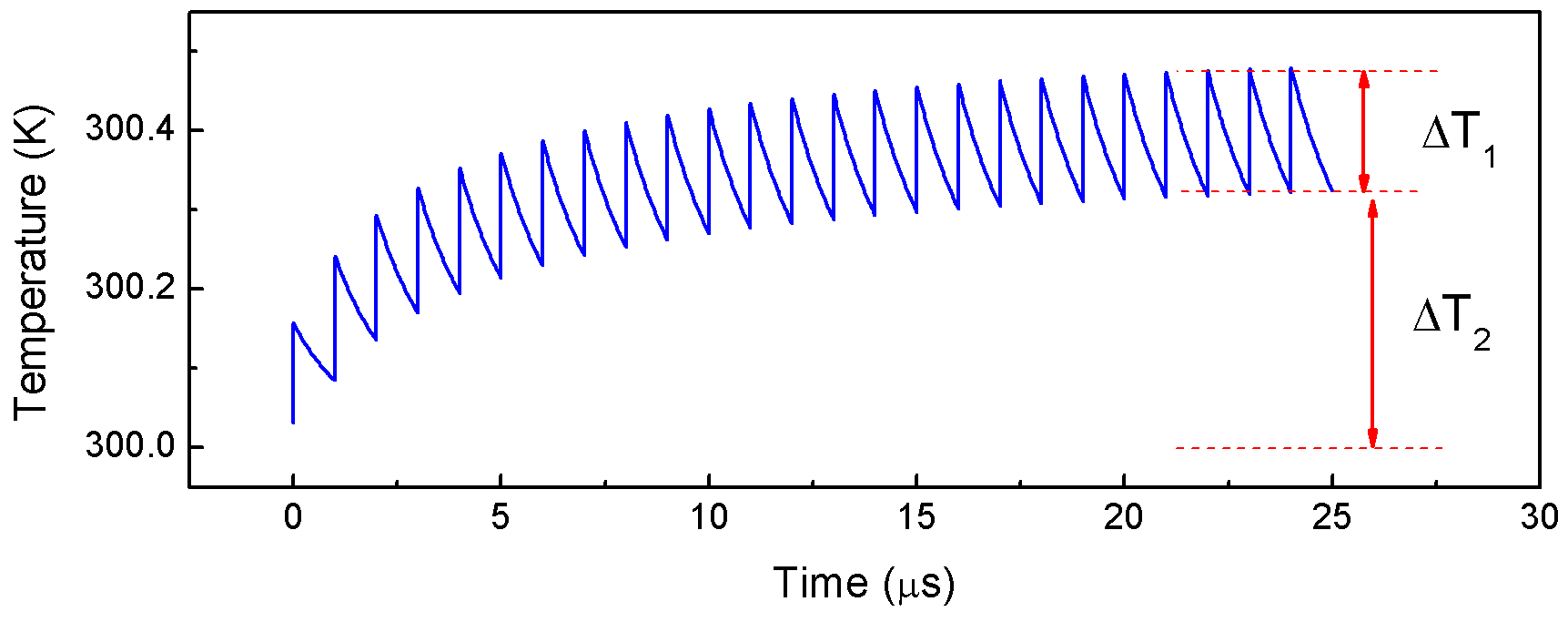}
\figcaption{\label{temperature300k}   Temperature changes at the crystal center under 300 K environmental temperature. }
\end{center}

\section{Results and discussions}
Both user applications and the FEL buildup of an XFELO require spectral width less than 1 meV for the X-ray pulse \cite{kwang}. However, with increasing temperature, the crystal volume expands as well as the lattice spacing, which leads to a shift of the Bragg energy and broadening of the XFELO bandwidth. According to the formula $\triangle E/E=\beta \triangle T$ \cite{lab10,stoupin,kkm}, there exists a critical temperature change $\triangle T_c$ for diamond crystal to maintain the required X-ray bandwidth ($<$1 meV). As shown in Fig.~\ref{temperature300k}, for XFELO operation, both $\triangle T_1$ from X-ray pulse energy deposition and  $\triangle T_2$ from remaining heat accumulation may degrade the buildup of XFELO from shot noise and the output stability after XFELO saturation. Thus a criterion for the total crystal temperature change $\triangle T_t$ can be written as follows:
\begin{equation}
\label{eqn3}
\triangle T_t=\triangle T_1+\triangle T_2<\triangle T_c
\end{equation}

In this section, the heating dependence on the environmental temperature, crystal size, X-ray pulse repetition rate and pulse energy are investigated.

\subsection{Environmental temperature}
To investigate the effect of environmental temperature, we used the baseline parameters for the X-ray and diamond crystal, while setting the environmental temperature as 77 K, 100 K, 200 K and 300 K respectively. After interaction with the X-ray pulse in a cold environment such as 77 K, shown in Fig.~\ref{temperature}, the crystal temperature rises to a higher level and recovers more quickly than that at the high temperature of 300 K (Fig.~\ref{temperature300k}).
\begin{center}
\includegraphics[width=8cm]{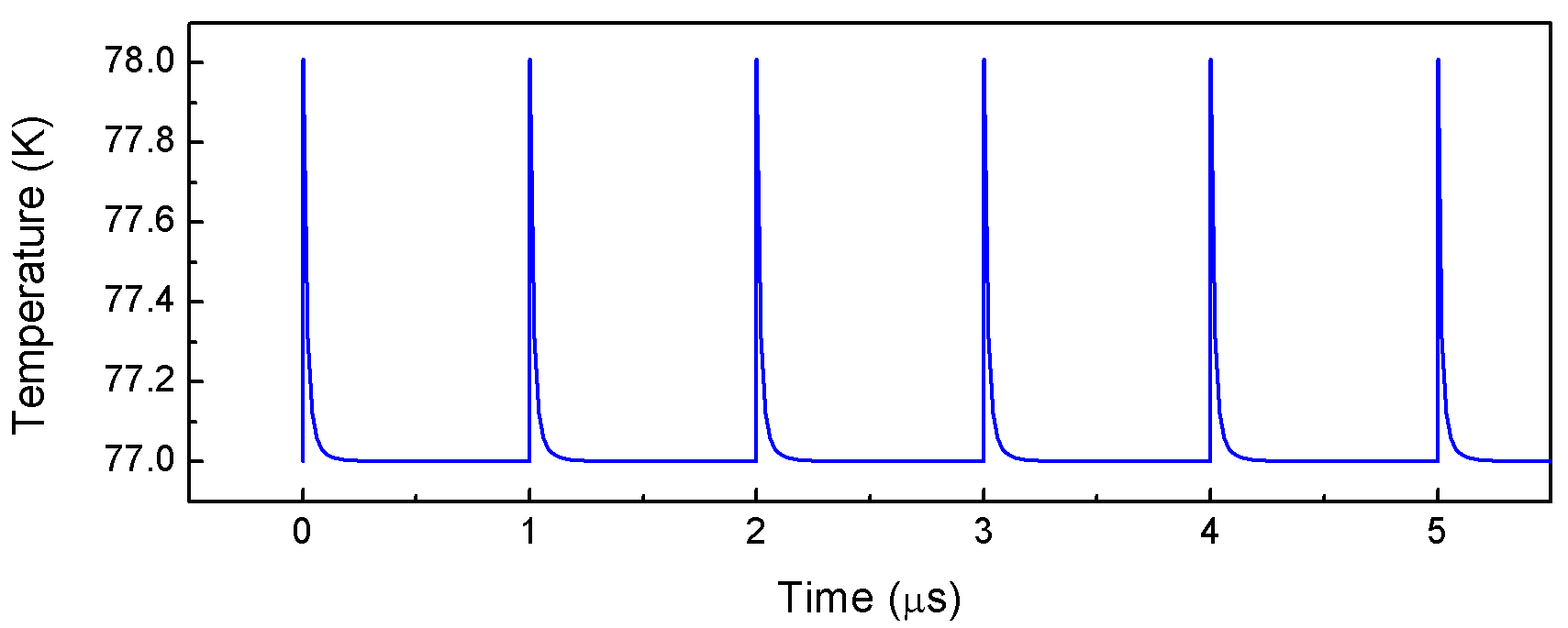}
\figcaption{\label{temperature}   Temperature changes at the crystal center under 77 K environmental temperature. }
\end{center}

The thermal conductivity $\lambda$, specific heat capacity C, and thermal expansion coefficient $\beta$ of diamond crystal varies dramatically as the external temperature changes \cite{lab17}, as summarized in Table~\ref{tab1}. Obviously, a lower environmental temperature ensures higher thermal conductivity coefficient of the diamond crystal, which guarantees heat is lost quickly and the crystal temperature is almost recovered before the arrival of the next X-ray pulse.
\begin{center}
\tabcaption{ \label{tab1}  The thermal parameters and tolerated temperature shift of diamond crystal with varying environmental temperature. \cite{lab17}}
\footnotesize
\begin{tabular*}{80mm}{c@{\extracolsep{\fill}}lcccr}
\toprule T/K & 77   & 100  & 200 & 300 \\
\hline
$\lambda$ /$W(m\cdot K)^{-1}$ \hphantom{00} & \hphantom{0}8973 & \hphantom{0}7845 & \hphantom{0} 4129 & 2347  \\
C /$J(kg\cdot K)^{-1}$ \hphantom{00} & \hphantom{0}7 & \hphantom{0}15 & \hphantom{0} 122 & 410  \\
$\beta$/$K^{-1}\times 10^{-8}$ \hphantom{00} & \hphantom{0}$3.4$ & \hphantom{0}$7.4$ & \hphantom{0} $60$ & $120$  \\
\bottomrule
\end{tabular*}
\end{center}

Fig.~\ref{temperaturechange} illustrates the crystal temperature changes at different environmental temperatures. As the external temperature increases, $\triangle T_1$ decreases and $\triangle T_2$ increases because of independent behavior of special heat capacity and thermal conductivity, respectively. $\triangle T_t$ then shows a monotonic decrease. For the baseline case discussed here, when the environmental temperature is 77 K, 100 K and 200 K, the total temperature shift $\triangle T_t$ is lower than $\triangle T_c$. However, $\triangle T_t$ at 300 K is higher than its corresponding $\triangle T_c$, which means it is not acceptable.
\begin{center}
\includegraphics[width=7.5cm]{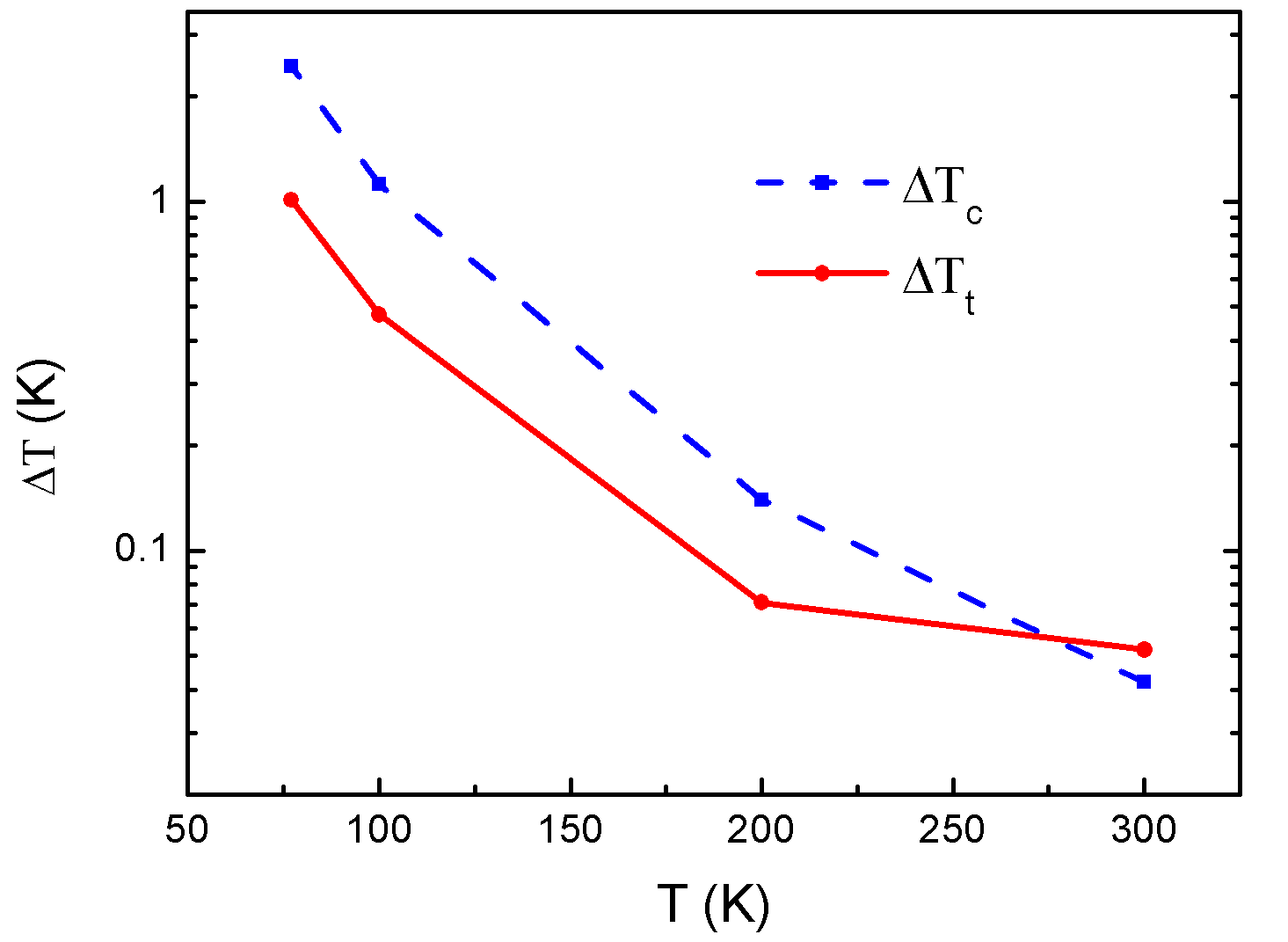}
\end{center}
\begin{center}
\includegraphics[width=8cm]{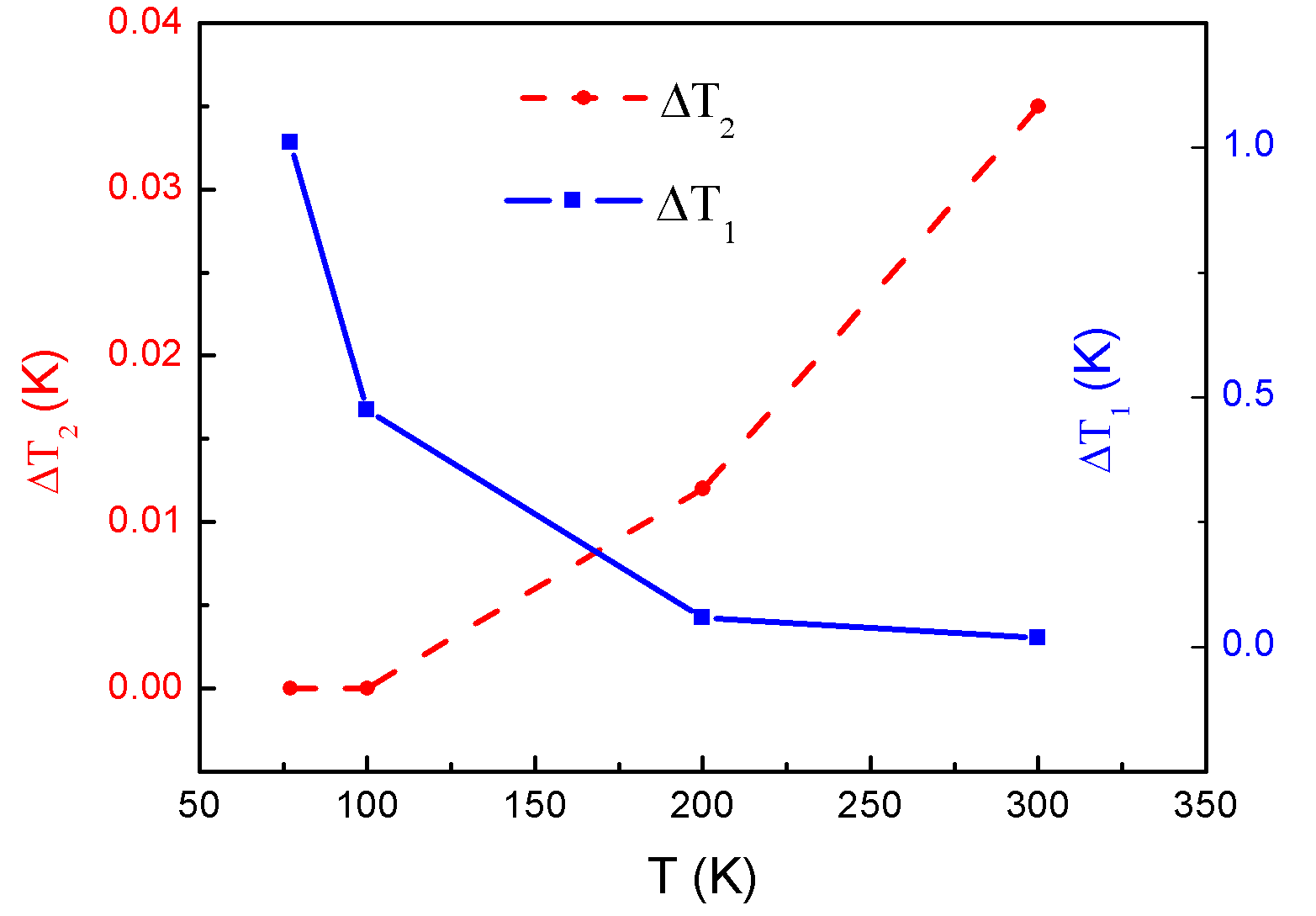}
\figcaption{\label{temperaturechange}   Crystal temperature change with varying environmental temperature. }
\end{center}

\subsection{Crystal size}
The size of diamond crystal has two aspects: transverse cross sectional area and thickness. To begin with, we study the relation between crystal transverse cross section area and temperature change at the center of crystal. The deposited heat is conducted outside mainly through the surface of a thin crystal, thus it is obvious that as the transverse area increases, the temperature shift is nearly constant as long as the cross section area is large enough to cover most of the incident X-ray photons. It agrees well with simulation results.

The influence of crystal thickness was studied by applying a heat source following the power law distribution Eq.~(\ref{eqn2}) to ANSYS analysis. Taking different thicknesses of crystal in ANSYS, the simulation results are shown in Fig.~\ref{thickness}. It is found that not only can a thinner mirror cool down faster, but it can also achieve a lower stable temperature. The reason is that the thinner the crystal is, the quicker it conducts heat to the environment.

\begin{center}
\includegraphics[width=8cm]{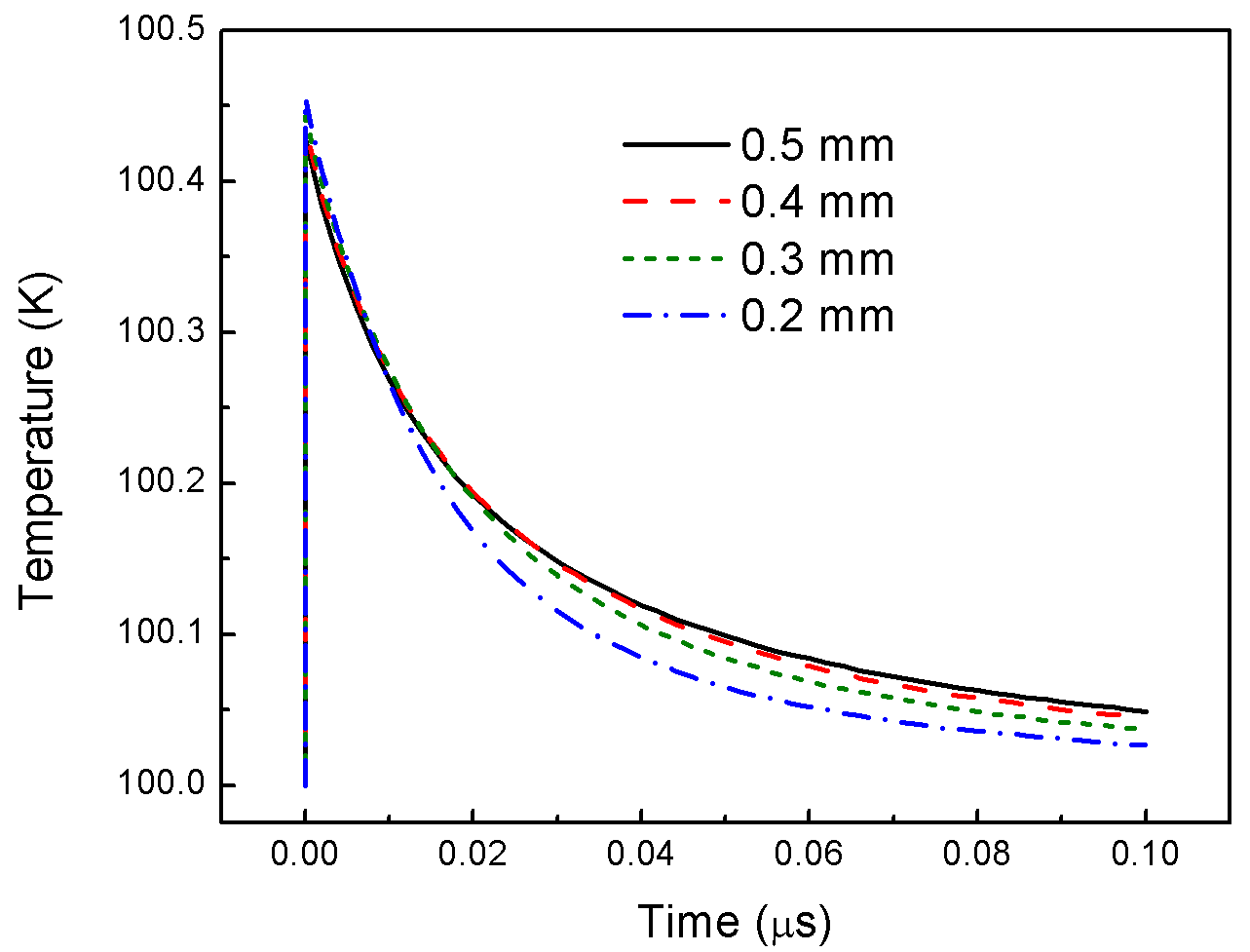}
\figcaption{\label{thickness}   The highest temperature change during cooling, with varying crystal thickness.}
\end{center}

For a 0.5 mm thickness crystal, the temperature evolution on the longitudinal axis during the cooling process is shown in Fig.~\ref{thickness2}. As time goes by, the highest temperature location moves towards the center of the crystal. The crystal thickness has less influence on XFELO than the environmental temperature.
\begin{center}
\includegraphics[width=8cm]{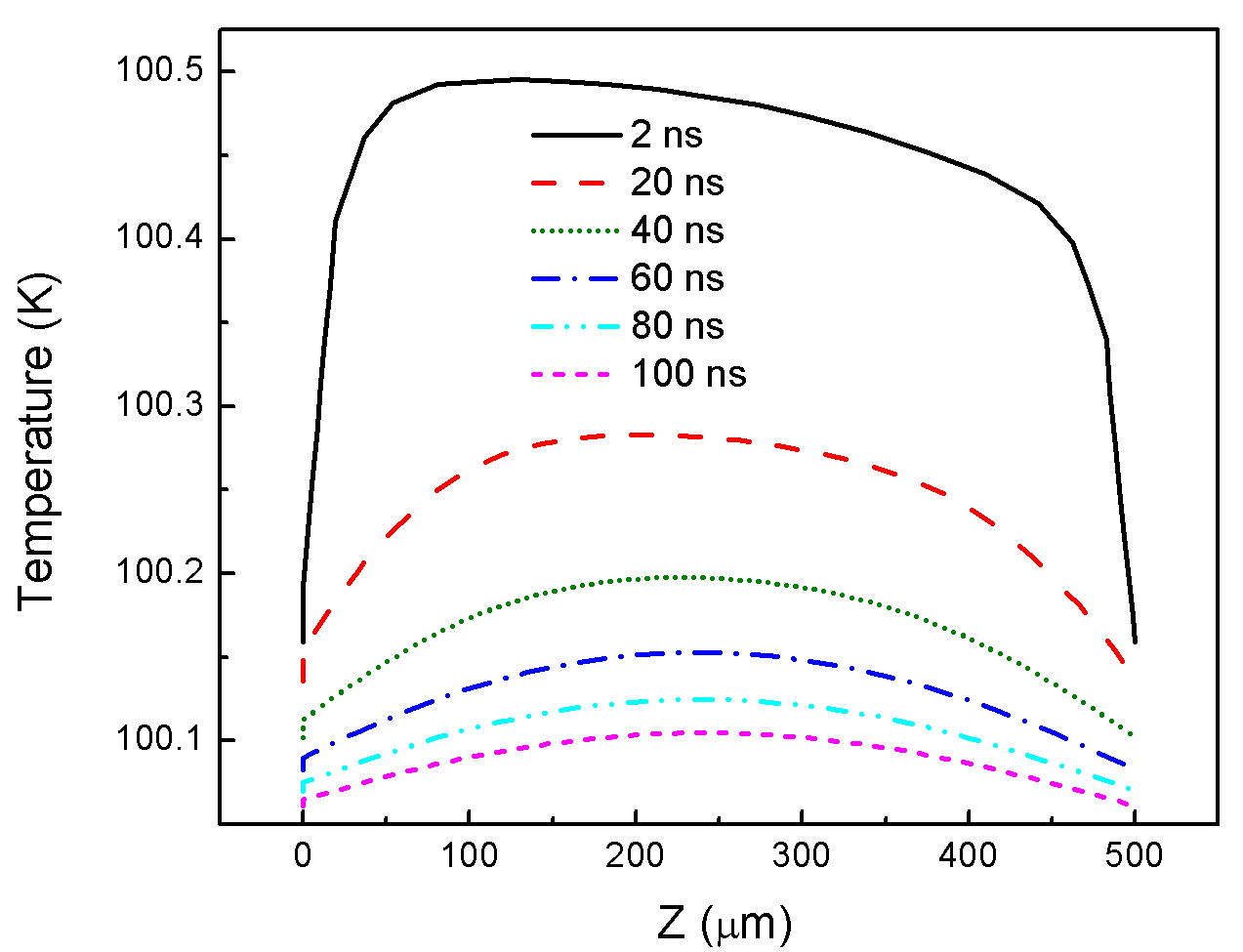}
\figcaption{\label{thickness2} Temperature change on the crystal longitudinal axis, with thickness of 0.5 mm and environmental temperature of 100 K.}
\end{center}

\subsection{X-ray pulse repetition rate}
To study the impact of the incident X-ray pulse repetition rate on the crystal temperature, we gradually increased the repetition rate from 1 MHz to 80 MHz for the 100 K case. Maximum temperature as a function of X-ray repetition rate is shown in Fig.~\ref{frequency}. With the repetition rate increase, the time available for cooling is shortened, which means that the crystal temperature cannot fully recover before the arrival of the next X-ray pulse, thus the temperature change $\triangle T_2$ will rise because of residual heating accumulation. As mentioned in previous section, in order to keep the bandwidth less than 1 meV, the highest temperature should be below 101.12 K. Combined with the simulation results, the acceptable highest pulse repetition rate is 63.64 MHz for the X-ray photon pulses assumed in Section 2.
\begin{center}
\includegraphics[width=8cm]{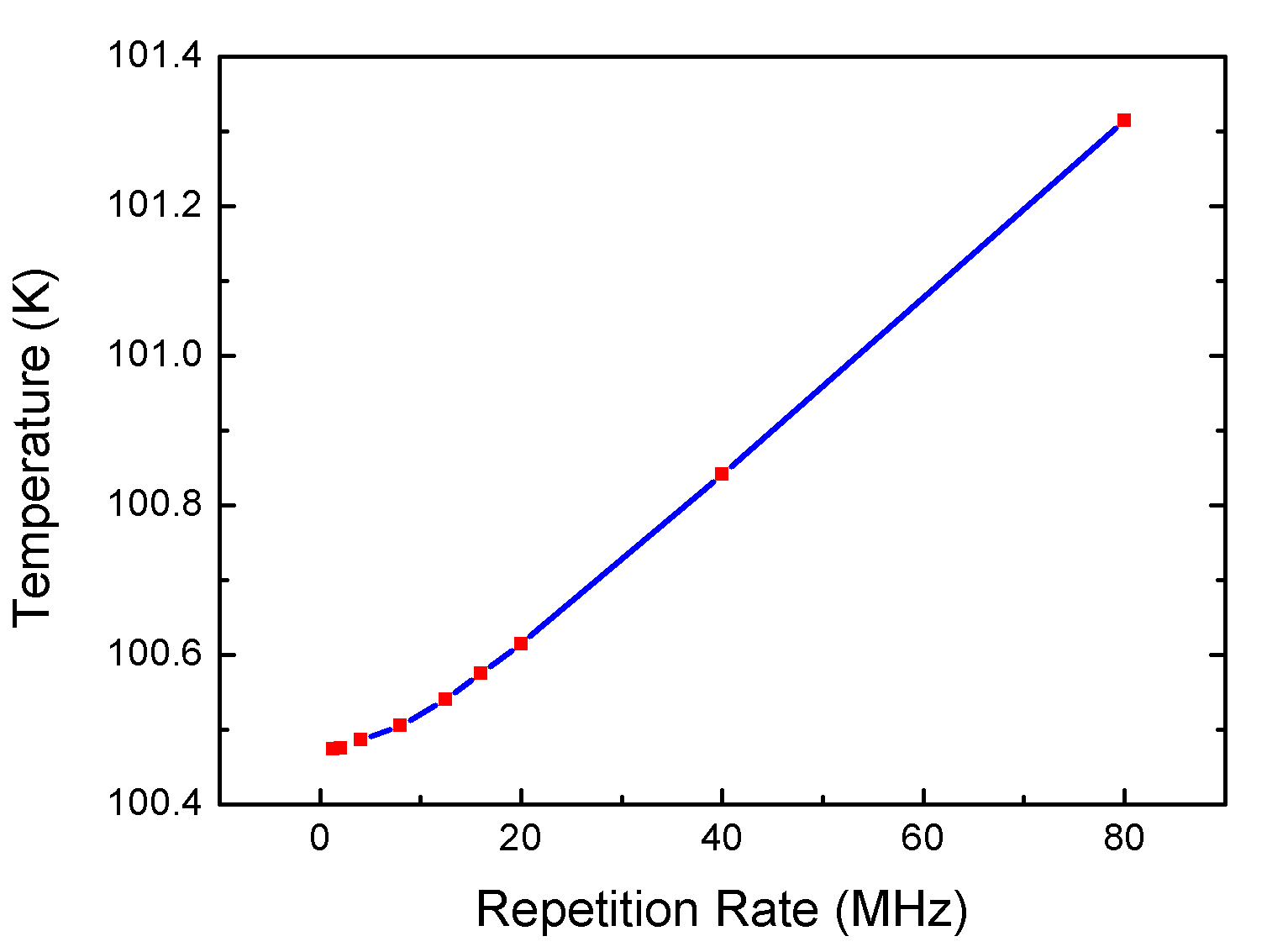}
\figcaption{\label{frequency}  The highest crystal temperature with different X-ray pulse repetition rate. }
\end{center}

\subsection{X-ray pulse energy}
As mentioned above, the thermal loading of a X-ray photon pulse is a function of photon flux and distribution, crystal reflectivity and thickness. It performs as a source providing heating energy to raise the temperature. If 1$\times 10^{10}$  photons are injected into a 200 $\mu$m thick crystal, 0.15 $\mu$J energy is absorbed under the conditions we have chosen. Because the X-ray pulses are so short and the temperature increase is so small ($<$5 K) that thermal conduction during the heating process and heat capacity shift can be ignored, the highest crystal temperature is proportional to the incident X-ray pulse energy, as illustrated in Fig.~\ref{energy}. To satisfy the energy bandwidth criterion ($\triangle E <$ 1 meV), the X-ray pulse energy should be less than 45 $\mu$J, under the environmental temperature of 100 K. It is worth stressing here that the X-ray pulse energy in the cavity is 43.5 $\mu$J for the original XFELO proposal operated at the fundamental of the resonant wavelength \cite{lab9}, while the typical photon pulse energy in the cavity is 3.4 $\mu$J for a harmonic lasing scheme XFELO \cite{lab12}, which is proposed to cut down the size and cost of large-scale machines.
\begin{center}
\includegraphics[width=8cm]{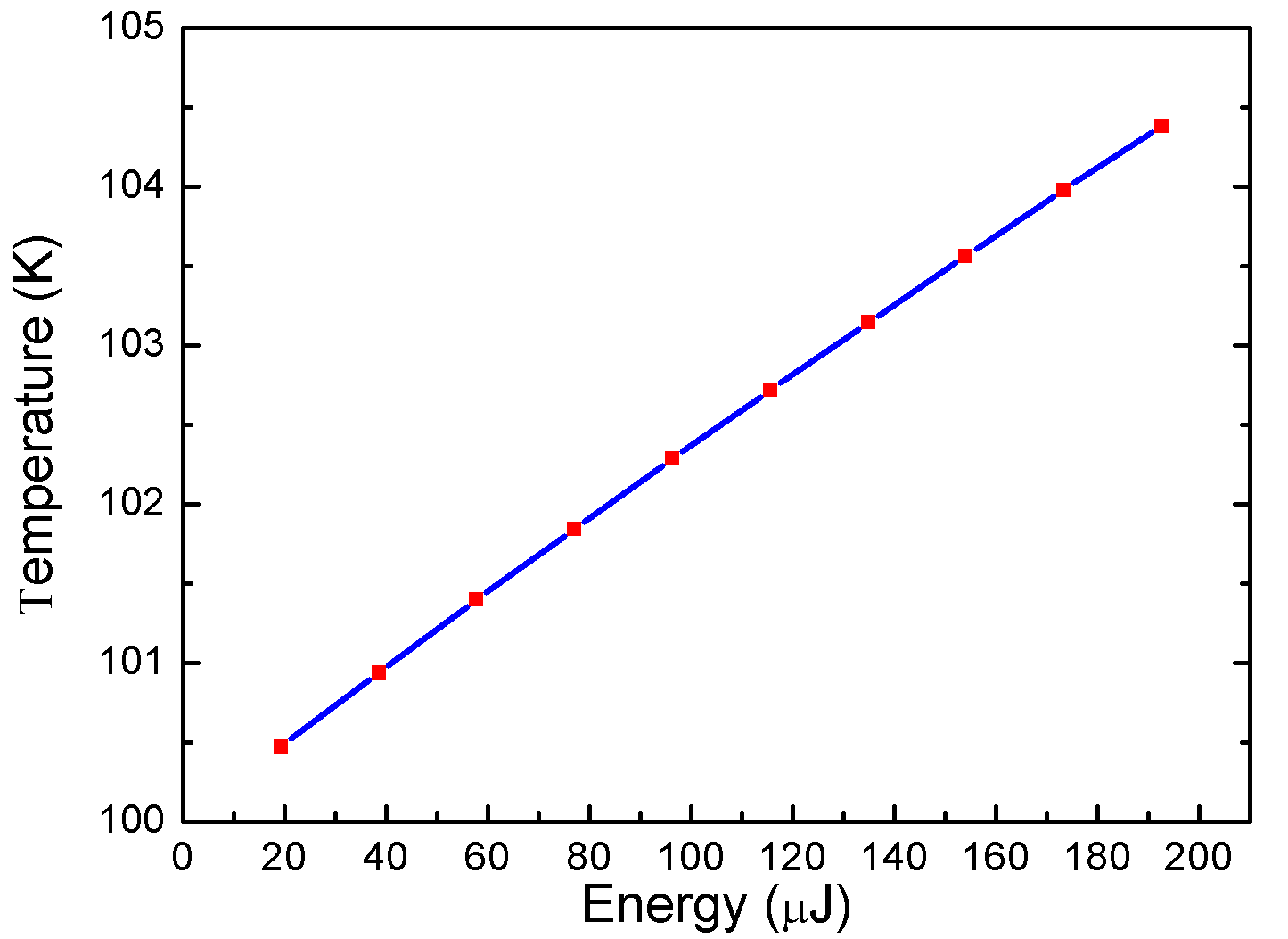}
\figcaption{\label{energy}  The highest crystal temperature with different X-ray pulse energy. }
\end{center}

\section{Conclusions}
Diamond crystal is commonly accepted as the material for fabricating mirrors for X-ray FEL oscillators. One of the most critical questions still open is how the thermal loading of diamond crystal degrades the performance of the X-ray source. In this paper, Geant4 and ANSYS are jointly used to simulate the interaction between the X-ray photon pulse and diamond crystal, and the crystal temperature shift due to thermal loading. Our results indicate that a diamond crystal with large transverse cross section and small thickness, and a low environmental temperature, are helpful to reduce the thermal loading effects in XFELO. For the XFELO example operated at 1 MHz repetition rate in this paper, in order to ensure the bandwidth of 1 meV, the X-ray pulse energy in the Bragg cavity should be lower than 45 $\mu$J under 100 K environmental temperature.

It is expected that this result is useful for the manufacture and establishment of a feasible crystal mirror in X-ray FEL oscillators. It is worth stressing that this study is preliminary and there are still several practical physical effects that are not included, such as accurate calculation of crystal reflectivity from dynamic theory \cite{lab18}, the noisy start-up and coherent build-up of XFELO, and the thermal expansion of the crystal lattice. These and other effects will be left for our subsequent reports. The numerical method used here can be easily extended to model the mono$-$chromator in hard X-ray FEL self-seeding schemes.

\vspace{10mm}

\acknowledgments{The authors would like to thank Shuaishuai Shen, Bo Liu and Dong Wang for helpful discussions.}

\end{multicols}

\vspace{-1mm}
\centerline{\rule{80mm}{0.1pt}}
\vspace{2mm}

\begin{multicols}{2}

\end{multicols}

\clearpage

\end{document}